\documentclass[letter]{aa} 

\usepackage{graphicx}
\usepackage{txfonts}

\newcommand{\erg}{${\rm erg \ s^{-1}}$ }

\def\ltsima{$\; \buildrel < \over \sim \;$}
\def\simlt{\lower.5ex\hbox{\ltsima}}
\def\gtsima{$\; \buildrel > \over \sim \;$}
\def\simgt{\lower.5ex\hbox{\gtsima}}
\newcommand{\msun}{{\rm\,M$_\odot$}}

\newcommand{\srcs}{{\rm\,J1302}}
\newcommand{\src}{{\rm\,J1302 }}

\newcommand{\xmm}{{\it XMM-Newton} }

\begin{document}

   \title{
Possible $\sim$0.4 hour X-ray quasi-periodicity from an ultrasoft active galactic nucleus
}

   \author{J. R.~Song
          \inst{1}
          \and
          X. W.~Shu
           \inst{1}
           \and 
         L. M.~Sun
           \inst{1}
           \and
        Y. Q.~Xue
           \inst{2} 
           \and  
        C.~Jin
           \inst{3, 4} 
           \and
        W. J.~Zhang 
           \inst{1}
           \and
        N.~Jiang
             \inst{2} 
          \and
        L. M.~Dou
           \inst{5}
          \and
          T. G.~Wang
           \inst{2}        
          }

   \institute{Department of Physics, Anhui Normal University, Wuhu, Anhui 241000, China \\
   \email{xwshu@ahnu.edu.cn} 
         \and
    CAS Key Laboratory for Researches in Galaxies and Cosmology, Department of Astronomy, University of Science and Technology of China, Hefei, Anhui 230026, China
         \and
    National Astronomical Observatories, Chinese Academy of Sciences, Beijing 100101, China 
         \and 
School of Astronomy and Space Sciences, University of Chinese Academy of Sciences, Beijing 100049, China
         \and
         {   Department of Astronomy,} Guangzhou University, Guangzhou 510006, China                   
             }


 
  \abstract
   {
 RX J1301.9+2747 is an ultrasoft active galactic nucleus (AGN) with unusual X-ray variability that is 
 characterized by a long quiescent state and a short-lived flare state. The X-ray flares are found to recur 
 quasi-periodically on a timescale of 13--20 ks. 
 Here, we report the analysis of the light curve in the 
 quiescent state from two \xmm observations spanning 18.5 years, along with the discovery of a possible 
 quasi-periodic X-ray oscillation (QPO) with a period of $\sim$1500s. The QPO is detected at the same 
 frequency in the two independent observations, with 
 {a combined significance of $>$99.89\%. }
 The QPO is in agreement with the relation between frequency and black hole mass ($M_{\rm BH}$) that has been reported in previous works for AGNs 
 and Galactic black hole X-ray binaries (XRBs). 
 The QPO frequency is stable over {almost two decades}, suggesting that it may correspond to the high-frequency type found in XRBs and originates, perhaps, from a certain disk resonance mode. 
 In the 3:2 twin-frequency resonance model, 
 our best estimate on the $M_{\rm BH}$ range implies {that a maximal black hole spin can be ruled out}. 
 We find that all ultrasoft AGNs reported so far display quasi-periodicities in the X-ray emission, 
 suggesting a possible link on the part of the extreme variability phenomenon to the ultrasoft X-ray component. 
 This indicates that ultrasoft AGNs could be the most promising candidates in future searches for X-ray periodicities.  
 }

  
   {}
   {}
   {}

   \keywords{Galaxies: active--
             X-rays: individual (RX J1301.9+2747)--
             Black hole physics
               }

   \maketitle
   %

\section{Introduction}

Active galactic nuclei (AGNs) are powered by the accretion of matter onto supermassive black holes (SMBHs)
with $M_{\rm BH}\sim10^6-10^9$\msun. They are considered to be scaled-up counterparts of Galactic 
black hole X-ray binaries \citep[XRBs, $M_{\rm BH}\sim10$\msun,][and references therein]{McHardy2006}. 
One example of the similarities between these two types of systems is the rapid X-ray 
variability \citep[e.g.,][]{Gierliski2008a, Middleton2010, Zhou2015}. 
High-frequency quasi-periodic oscillations (HFQPOs; 40--450 Hz) have been 
detected in the power spectra of XRBs at their very high accretion states \citep[see, e.g.,][for a review]{Remillard2006}.
As the fastest coherent features in the X-ray emission, these HFQPOs are 
expected to originate from the innermost regions of the accretion flows \citep[e.g., ][]{Lai2009}. 
In addition, the frequencies of HFQPOs appear stable and do not vary 
significantly with the X-ray luminosity, 
suggesting that {this might be a basic property that could }potentially provide a probe 
into the BH mass and spin \citep[][]{Abramowicz2001, Abramowicz2012, Motta2014, Sramkova2015, Goluchov2019}. 
 

  If the accretion process is dominated by {strong gravity and is independent }of BH masses, 
  a scale invariance implies that QPOs should also be present in AGNs. 
  RE J1034+36 is the first AGN for which a significant QPO has been detected \citep[][]{Gierliski2008b}. 
  However, a systematic study of power spectra of 104 AGNs by \citet[][]{Gonzalez2012} resulted in no 
  conclusive detections of QPOs, except for RE J1034+36.  
  {Possible detections of QPOs ($f_{\rm QPO}\sim[0.7-2.7]\times10^{-4}$Hz) were reported in a few other AGNs from dedicated 
  searches, including 2XMM J123103.2+110648 \citep[J1231,][]{Lin2013},} MS2254.9-3712 \citep[][]{Alston2015}, 
  1H 0707-495 \citep[][]{Pan2016, Zhang2018}, Mrk 766 \citep[][]{Zhang2017}, 
  and MCG -6-30-15 \citep[][]{Gupta2018}. 
  In addition, QPOs were also detected in the stellar tidal disruption events (TDEs) by SMBHs 
  where transient accretion is triggered \citep[][]{Reis2012, Pasham2019}.
  
It has been suggested that X-ray QPOs tend to occur in the highest accretion state 
  \citep[][]{Remillard2006, Pan2016}. 
   {Furthermore, by analyzing the latest X-ray observational data for RE J1034+36, \citet[][]{Jin2020a}
   confirmed the presence of QPO signal with an even higher significance. Interestingly, 
   they found that the QPO in 0.3--1 keV is leading by 1--4 keV and that  the QPO   
   is only detected in the spectral state, where the soft X-ray excess emission is relatively weak \citep[see also][]{Middleton2011}. 
   The hard X-ray power-law component, however, remains unchanged among the observations with and without the QPO.   
   These results suggest that the QPO in RE J1034+36 is linked to the soft X-ray spectral component, which possibly 
   originates from the inner disk \citep[][]{Jin2020b}. 
   }
   
    \begin{figure*}
           \centering
           \includegraphics[width=0.99\textwidth]{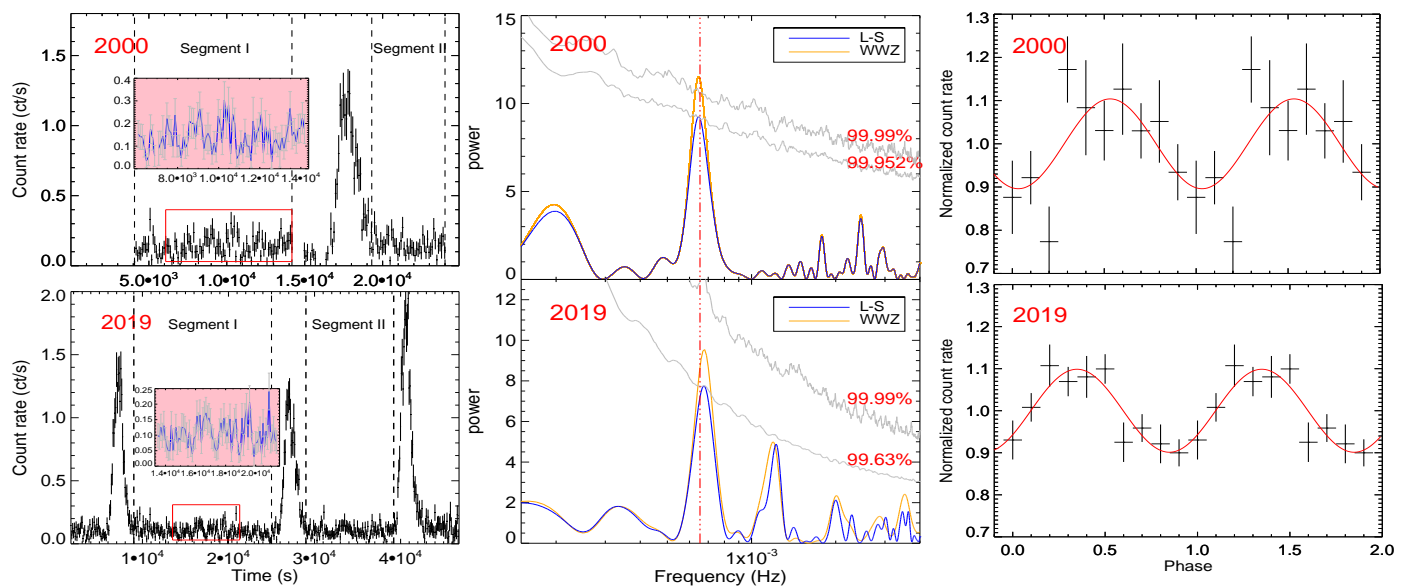}
           \caption{{\it Left:} X-ray light curves of \src in the energy range of 0.2-2.0 keV. 
           The upper panel is for the XMM2000 observation, while the lower is for the XMM2019 observation, respectively. 
           The inset panel indicates the segment used for the light curve analysis in this paper.
           {\it Middle}: Lomb-Scargle power as a function of frequency is shown in blue, while the time-averaged WWZ power 
           is shown in orange. Strong peaks are seen at a frequency of $ \sim 6.7 \times 10^{-4} $ Hz ($\sim$1500 s) which is indicated with 
           the vertical dash-dotted line. {The grey solid lines are single-trial significance curves obtained from the light curve simulations, 
           calculated at each individual frequency bin (local significance, see text for the details). }
           {\it Right}: Folded X-ray light curve with a period of 1500s (two cycles are shown). The best-fit sinusoid is shown in 
           red solid line. 
            } 
                 \label{FigGam}%
                 \vspace{0.2cm}
       \end{figure*}
       \vspace{0.6cm}

   \newpage

   
   {Among the AGNs with QPO detections, 
   J1231 is peculiar} in terms of its ultrasoft X-ray spectra, which lack significant emission at energies above $\sim$2 keV 
   and can be described entirely by a soft thermal component with a temperature of $kT$$\sim$0.12 keV \citep[][]{Terashima2012}. 
   Such an extreme soft emission is 
   unprecedented among AGNs. A similar level of extremely soft X-ray spectrum was also found in GSN 069 by \citet[][]{Miniutti2013}. 
   In either case, the thermal X-ray spectrum appears to be a close analog to the disk-dominated spectrum that is typically seen in 
   the high and soft states of XRBs. However, it was claimed that the ultrasoft X-ray spectra in the two AGNs could be associated with 
   TDEs because both clearly show a long-term decline in the X-ray flux \citep[][]{Lin2017, Shu2018}. More interestingly, 
   \citet[][]{Miniutti2019} discovered a variability of $\sim$9 hour X-ray quasi-periodic eruptions (QPEs) in GSN 069 during its flux decay phase, 
   casting intriguing questions about the origin of its ultra-soft X-ray emission.

   Besides J1231 and GSN 069, the ultra-soft X-ray emission is also detected in RX J1301.9+2746 
   ({hereafter, J1302,} Sun, Shu \& Wang 2013, Sun13). J1302 is the first AGN known to display  
   short-lived X-ray flares over a stable quiescent flux state. 
   Recent \xmm observations confirmed the recurrence of the X-ray flares, with an amplitude similar to that of the QPEs in GSN 069 
   \citep[][]{Giustini2020}. 
    In this work, we report the detection of a possible X-ray QPO signal at 
    $[6.68\pm0.52]\times10^{-4}$ Hz (i.e., corresponding to 
    $\sim$0.4 hour) in 
    the light curve of the quiescent state for J1302. 
    The signal was detected independently in two \xmm observations, in 2000 December and 2019 May, respectively, 
    with a combined significance level {of $>$99.89\% based on a  consideration of the number of 
    individual frequencies over which the QPO was} investigated. 
    The QPO has the highest frequency ever seen from an AGN, which is possibly due to the low BH mass for the source. 
    Its remarkable stability {over almost two decades} may represent a fundamental property of the system. 
    In Section 2, we describe the observations and data reduction. In Section 3, we present the 
    light curve analysis and results. A discussion on the origin of the QPO and its implications are  
    presented in Section 4.  

\section{Observations and data reduction}




J1302 was observed twice by \xmm {with the European
   Photon Imaging Camera (EPIC)}, {in 2000 December for an exposure time of 29 ks 
        (OBSID: 0124710801, hereafter, XMM2000) 
and in 2019 May} for about 48 ks (OBSID: 0851180501, XMM2019 hereafter). 
   While the \xmm data have already been presented in Sun13 and \citet[][]{Giustini2020}, here, we re-analyze 
   the light curve in detail, especially at the quiescent state with the aim of searching for 
   periodicity in the X-ray emission. 
   The \xmm data were processed following the standard procedures, using the Science Analysis Software (SAS)
   version 17.0.0, with the latest calibration files as of 2018 December. 
   The epochs of high background events were examined and
   excluded using the light curves in the energy band above
   12 keV. 
   Source events were extracted from a circular region with a radius of 40\arcsec~centered 
   at the source position, while background events were from source-free areas on the same chip 
   using four circular regions identical to the source region. 
   We primarily used  the data from the EPIC PN camera, which have a much
   higher sensitivity. 
   Only good events (i.e., PATTERN$\le$4 for PN)
   were used to generate light curves. 
   The SAS tool {\it epiclccorr} was used to correct for 
   instrumental effects. Given the low count rates in the quiescent state ($\sim0.1$ counts/s), 
   the photon pile-up effect is negligible. 
   
 \section{Light curve analysis and results}
   
   Figure 1 (left panel) shows the background-subtracted light curves that were constructed in the 0.2--2 keV band   
   with a bin size of 100s, based on the XMM2000 and XMM2019 observations, respectively. 
   Apart from the flares that were reported in previous works, 
   a periodicity appears in the X-ray emission during the quiescent state. 
   {Timing studies of XRBs have suggested that QPOs can evolve with time in their amplitudes or frequencies, or both \citep[][]{Remillard2006}. 
   Such a time dependence has also been found in AGNs where the QPO signal is only seen in segment of light curves \citep[e.g.,][]{Gierliski2008b, Pan2016, Zhang2017}. 
   Weighted Wavelet Z-transform (WWZ) is a powerful technique for analyzing and measuring non-stationary periodic features 
   of the time series by performing signal decompositions in both frequency and time spaces simultaneously 
   \citep[][]{Foster1996, Czerny2010}. 
   We calculated the WWZ power as a function of frequency and time for the segment of quiescent light curve  
   in the range of $t=4000-14100$s for the XMM2000 observation and $t=9000-25000$s for 
   the XMM2019 observation, respectively (segment I in Figure 1). 
   The contour plots of the WWZ power in the time-frequency plane 
   for both observations (Figure 1 in the appendix) suggest that a potential 
   QPO signal is present within a frequency range of $\sim6\times10^{-4}-8\times10^{-4}$Hz, 
   but does not appear throughout the whole light curve. 
   Therefore, only the segment where the signal is relatively strong was selected for the 
   following time analysis, which is $t=6000-14100$s for the XMM2000 data, and $13500-21300$s for the XMM2019 data, respectively. 
   {We performed} similar WWZ analyses on the other segments of the light curve (Segment II), but 
   did not find any significant signal at the similar frequency (Figure 2 in the appendix). 
   Therefore, these segments are not considered further in our time analysis. 
   

   In Figure 1 (middle panel), we plot the time-averaged WWZ power as a function of frequency (orange lines). 
   The power spectra of both the XMM2000 and XMM2019 observations display a strong peak near $6.7\times10^{-4}$ Hz,
   or, potentially, a QPO component. }
   By fitting a Gaussian function to the signal, we obtained a centroid frequency ($\nu$) of $6.54\times10^{-4}$ Hz and 
   $6.82\times10^{-4}$ Hz, with frequency width (full-width at half-maximum) of $\delta\nu=1.11\times10^{-4}$ Hz and 
   $\delta\nu=1.30\times10^{-4}$ Hz, for XMM2000 and XMM2019 observations, respectively. 
   This corresponds to a quality factor of $Q$=$\nu/\delta\nu\simgt5$, indicating a coherent feature.  
   We note that the centroid frequency of the signal for the XMM2019 observation is slightly higher 
   than that in XMM2000, 
   but it is not statistically significant within the errors. 
   Therefore, the average frequency from the two observations is reported as the QPO frequency, $f_{\rm QPO}=[6.68\pm0.52]\times10^{-4}$ Hz, 
   corresponding to a period of $\sim$1500s ($\sim$0.4 hour).

   {To further investigate and quantify the periodicity, we employed the Lomb-Scargle 
           Periodogram \citep[LSP,][]{Lomb1976, Scargle1982}, 
           which is a widely used method in the power spectrum analysis 
   of non-uniformly sampled time-series data.  
   This may be the case for our data, as the light curves may not be strictly evenly sampled 
   due to the effect of high-background intervals. 
   The LSP analysis for the above selected segments of light curve (inset panels in Figure 1) 
   provides very similar results to WWZ and 
   detects the QPO at the same centroid frequency (blue lines), albeit with a smaller amplitude of the peak power. }
   The significance of the LSP power can be assessed by testing the null hypothesis of no period being present.  
   It is also known as the false-alarm probability of obtaining a power larger than the measured peak, assuming that 
   the signal is on top of purely Gaussian-noise time series.  
   The false-alarm probability is $p=0.005$ and $p=0.03$, corresponding to a {global} confidence level of 
  99.5\% for XMM2000 and 97\% for XMM2019, respectively. 
   We note that this test and quoted significance rely on only white noise variability. 
   The presence of red noise is known to be a potential source of false features in the X-ray 
   power spectra of AGNs \citep[e.g.,][]{Vaughan2005}. 

   In order to rigorously estimate the confidence level, we employed Monte Carlo techniques 
   and compared the spread of a series of simulated power spectra to the observed one in a similar 
   manner to that outlined in \citet[][]{Uttley2002}. 
Firstly, we fitted the power density spectrum (PDS) of the observed light curve with a simple 
power-law function consisting of a power-law plus a constant:  
$ P (f )=N f^{-\alpha}+ C $ , where $ N $ is the normalisation term, $\alpha$ is the power-law index, 
and $C$ is a non-negative constant representing the Poisson noise level {(see Figure 3 in the appendix)}. 
A maximum likelihood method was used to determine the parameters in this model \citep[][]{Vaughan2010}. 
We obtained normalization $N=0.1245$, $\alpha=0.4647$, and $C$=${14.166}$. 
Simulated light curves for this best-fit power-law model were then constructed using the method of
\citet[][]{Timmer1995}, which were resampled and binned to 
have the same duration, mean count rate, and variance as the observed light curve.  
We generated their power spectra using the same Lomb-Scargle method. 
For each simulated power spectrum, we found the spread in the value of normalized LSP power 
and obtained the spurious detection probability 
that a value would be at or greater 
than the power of interest at a frequency bin. 
At the frequency of $f=6.68\times10^{-4}$ Hz, where the QPO is indicated, we found that there 
are only 48 data points out of $10^5$ with powers above the measured peak, 
which gives a confidence level of 99.952\% for the XMM2000 data. 
From the simulated light curves, we calculated the LSP power at each frequency  
corresponding to the confidence level of 99.952\% and constructed 
the significance curve that is shown in Figure 1 (middle panel, grey curve). 
For comparison, we also built the 99.99\% confidence curve following the same procedures.  
The same light curve simulations were performed for the XMM2019 data, which yield 
a significance of 99.63\% for the QPO signal. 
We note that using a bending power model to fit the PDS and construct 
simulated light curves results in similar confidence levels for the QPO detections.

The statistical tests above yield a probability of finding a QPO with 
the power {exceeding that found in the observed data in a given (predefined) frequency bin.  
In general, a QPO could have been} detected at any plausible frequency for a pure noise {power spectrum distribution (PSD)}. 
Hence, we proceeded to calculate the significance of a QPO peak globally by scanning all frequency  
bins over the range of interest ($1.2\times10^{-4}-5.0\times10^{-3}$Hz, Figure 1). 
For each of the simulated LSP power spectra, the maximum power ($\zeta_{\rm max}$) was noted. 
Using the $10^5$ values of $\zeta_{\rm max}$, we computed the cumulative distribution of the 
probability to exceed a given $\zeta_{\rm max}$ (Figure 4 in the appendix). 
By comparing with the observed $\zeta_{\rm max,obs}$, we found that the significance for the 
QPO detection is reduced to $96.81$\% and $96.65$\% for the XMM2000 and XMM2019 data, respectively. 
However, we show here that the {QPO exists} in two independent observations with 
a similar power level at the same frequency.  
The combined blind-chance probability that these two QPO signals arise from random noise alone is 
$P_{\rm false}=1.068\times10^{-3}$, corresponding to a detection at the $99.89$\% level, 
still making it statistically significant enough to claim the presence of a QPO. 
We note that the QPO feature {is preserved} at the same frequency in the PSD of the periodic light curve segment   
if the periodogram is calculated as the modulus-squared of the discrete Fourier transform (DFT), 
with an [rms/mean]$^2$ normalization \citep[][]{Vaughan2010}, as shown in Figure 3 in the appendix. 
On the contrary, no QPO feature is seen in the PSD of the full light curve. 
This suggests that either {the QPO} is transient or the QPO phase is shifted due to the interruption by flare intervals. 

   \begin{figure*}
           \centering
           \includegraphics[width=0.93\textwidth]{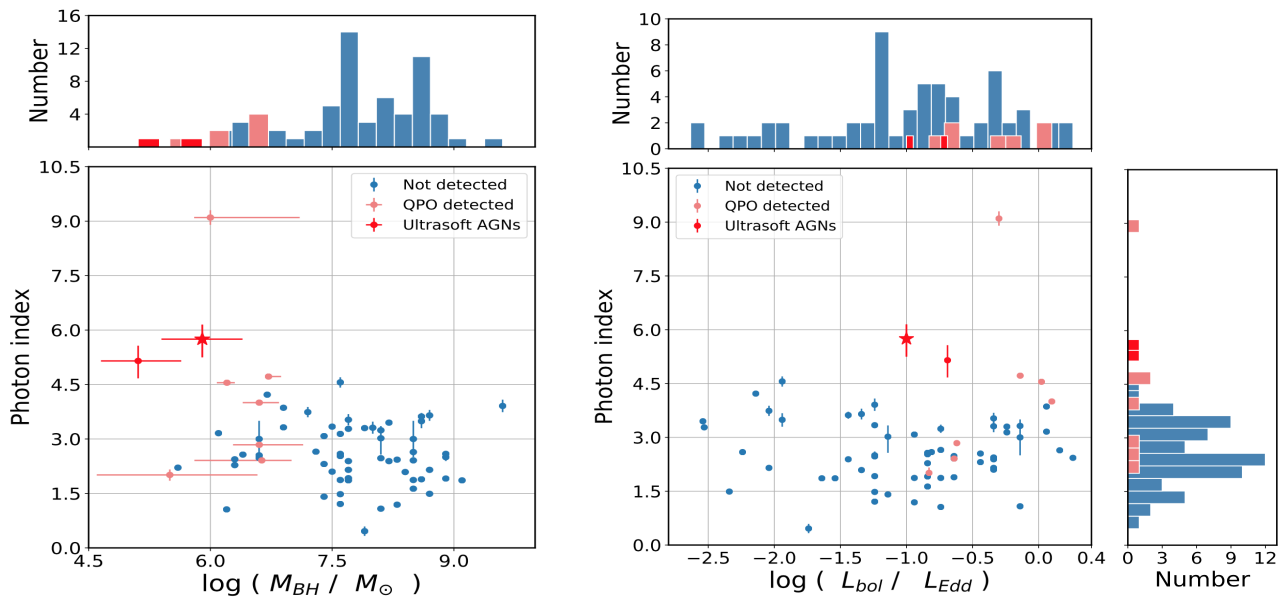}
           \caption{{{\it Left:} BH mass versus photon index for the SMBH accreting systems in which QPOs are detected, with 
            ultrasoft AGNs highlighted in red. The red star represents \srcs. The photon index was derived from the spectral fitting with 
            a simple absorbed power-law model to the soft X-ray spectrum below 2 keV. 
            For a comparison, we overplot the AGN sample of \citet[][]{Gonzalez2012} in which QPOs were searched uniformly (light blue), 
            but not found (except for RE J1034+36). 
            {Errors on the photon index are given by spectral fittings, corresponding to 90\% confidence level for 
                    one parameter of interest ($\Delta\chi^2=2.706$). 
                    The BH masses from the AGN sample of \citet[][]{Gonzalez2012} have characteristic 
                    errors of $\sim$0.5 dex \citep[][]{Vestergaard2006}.
            }
            {\it Right:} Same as left, but for the plot of Eddington ratio versus photon index. 
     Errors on the Eddington ratios are not shown due to the large uncertainties inherent in determining bolometric luminosity 
     and BH mass.
            Distributions of the BH mass and Eddington ratio and are shown on the top panel, while the distribution of the 
    photon index is shown on the right panel.} 
            }
                 \label{mdot}%
       \end{figure*}

With the light curve analysis tool {\tt efold} provided in the HEASOFT   software\footnote{https://heasarc.gsfc.nasa.gov/docs/xanadu/xronos/examples/efold.html}, 
we folded the light curves with the best-estimate of the period of
1500s for the two \xmm observations, which are displayed in the right panel of Figure 1 (two cycles are shown). 
The best-fit sinusoidal curve is overlaid in red. 
It can be seen that the amplitude of the X-ray flux clearly varies with phase. 
The fractional rms modulation amplitude derived from the folded light curve is 8.5$\pm1.3$\% 
and 6.8$\pm0.4$\% for the XMM2000 and XMM2019 data, respectively, 
{which is comparable to that in other AGNs. 
For example, the QPO rms variability for RE J1034+36 is found about 4\% in 0.3--1 keV and 12\% in the 1--4 keV \citep[][]{Jin2020a}.

}

\section{Discussion and conclusions}

The X-ray light curve for J1302 is peculiar among AGNs and characterized by two distinct states: a long
quiescent (or stable) state and a short-lived flare (or eruptive) state, which
differs in count rates by nearly an order of magnitude (Sun13). 
The X-ray flares are persistent and recurrent with quasi periods of $\sim13-20$ ks (Figure 1, see also 
Giustini et al. 2020), which are remarkably similar to the X-ray QPEs recently found in GSN 069 \citep[][]{Miniutti2019}.  
In this work, we have presented the timing analysis of light curves in the quiescent state with two different techniques (WWZ and LSP), 
which have revealed a potential QPO from the peak in the PDS, with a centroid frequency of $\sim6.7\times10^{-4}$Hz. 
{The QPO is present in two \xmm observations spanning 18.5 years, XMM2000 and XMM2019, at a confidence level of 96.81\% and 
96.65\%, respectively. 
Considering that the QPO was detected at the same frequency in two independent observations, 
its combined confidence level is 99.893\% ($\sim$3.27$\sigma$ if assuming normal distribution).} 
The relative stability of the QPO in frequency indicates its probable 
association with HFQPOs, as observed in XRBs.

Another peculiarity that characterizes \src is its ultrasoft X-ray spectrum in the quiescent state, where the typical AGN hard X-ray power-law emission 
above 2 keV is completely absent. The X-ray spectrum can be well described by a thermal disk emission, plus a much weaker 
Comptonization component \citep[][]{Shu2017}. 
While the disk emission appears stable between XMM2000 and XMM2019 observations, 
the additional harder spectral component has increased in flux by a factor of 3 \citep[][]{Giustini2020}. 
The spectral modeling suggests that the source could be in a disk-dominated thermal state, which, 
{considered analogously to XRBs, 
would suggest it is }accreting at high rates. The black hole mass of \src is not well-determined. 
Using the width of the {\sc [O~iii]}$\lambda$5007 as a
proxy for the stellar velocity dispersion of the host galaxy, we
obtained a BH mass of $M_{\rm BH}=8\times10^5$\msun, with an intrinsic
scatter of 0.5 dex. With a bolometric luminosity of $\sim10^{43}$\erg(Sun13), 
this implies an accretion rate in Eddington units of $L/L_{\rm EDD}\sim0.1$, suggesting that 
\src is indeed accreting at a high rate. 
These values are also consistent with the BH mass and Eddington ratio derived independently from 
detailed physical modeling of ultraviolet-to-X-ray energy spectra \citep[][]{Shu2017}.


High accretion rates are a characteristic feature of narrow-line Seyfert 1 galaxies 
\citep[NLS1s, e.g.,][]{Komossa2008, Boller2010}, 
which can have BH masses as low as that of \src \citep[][]{Zhou2006}. 
Interestingly, all five AGNs with potential HFQPO detections (including the prototype RE J1034+36) are NLS1s. 
The similar QPO frequency, X-ray spectral properties as well as the PDS shapes may suggest a common origin 
for the QPOs. 
The TDE Sw J1644+57 in which a 200-second QPO was detected is also believed to accrete at a super-Eddington rate \citep[][]{Reis2012}. 
This led Pan et al. (2006) to propose the possible link of QPOs to accretion rates. 
However, the QPO mechanism in \src may be different from those in either RE J1034+36 or Sw J1644+57, 
as {the QPOs for latter cases are more significant \citep{Gierliski2008b} or only shown }in hard X-rays 
\citep[$>$2keV,][]{Reis2012} and not in the soft state. 
 We note that by decomposing the soft excess emission of RE J1034+36 into multiple spectral components, 
including an inner disk emission and two warm Comptonized components, 
\citet[][]{Jin2020b} recently found that the QPO appears only in the warmer Comptonized component 
(hotter part of the soft excess) and in the disk component below 0.5 keV, with the latter leading 
by $\sim$680 s. 
This suggests that the QPO of RE J1034+36 may originate from the inner disk
and is then transmitted to the coronal regions via a Comptonization process. 
Detailed spectral and timing analysis of the soft excess emission in other NLS1s is required to 
test whether similar QPO components can be revealed from standard AGN stochastic variability, 
which is helpful to gaining an understanding of the origin of QPO. 
While a 3.8 hour QPO has been detected in the ultrasoft AGN J1231, 
it could be associated with a low-frequency QPO phenomenon (Lin et al. 2013).  
More recently, a remarkably stable QPO was detected in the TDE ASASSN-14li, which also has an ultrasoft 
X-ray spectrum \citep[][]{Pasham2019}, likely originating from a newly formed compact accretion disk. 
The results appear to be at odds with the scenario in XRBs suggesting the absence of QPOs in the soft (thermal) 
state \citep[][]{Remillard2006}; hence, this may require a new physical mechanism to explain the origin of QPOs 
in ultrasoft AGNs as well as TDEs.  

We fit a simple absorbed power-law model to describe the soft X-ray spectra for 
all of the SMBH accretion systems with QPO detections. 
The {resulting spectral indices plotted against the BH masses are shown in Figure 2 (left, red points). 
As a comparison, we overplot the AGN sample of \citet[][]{Gonzalez2012} in which 
a uniform search for QPOs was performed but yielded no detections. 
There appears to be a dependence of the detectability of QPOs on BH masses. 
About half of sources with QPO detections have a BH mass of $\simlt10^6$\msun, while only one AGN without QPO detection 
falls within this mass region. 
We also plot the soft X-ray spectral indices versus Eddington ratios, as shown in Figure 2 (right). 
It can be seen that while the QPOs tend to occur in the high-accretion systems, with $L/L_{\rm EDD}\simgt0.1$, 
most (6/9) have a steep soft X-ray spectrum with a photon index $\Gamma\simgt4$. 
In contrast, for AGNs in which no QPOs were detected, only few of them have the steep photon index as high as 
$\Gamma=4$. 
This suggests that the appearance of QPOs may be closely linked to the soft X-ray spectral component, 
especially in the disk-dominated state. 
}
Strictly speaking, the comparison with the AGN sample of \citet[][]{Gonzalez2012} 
may not be appropriate as the QPO was searched using the full light curves. 
This would reduce the significance of QPO detection if the signal is present in only part 
of light curve \citep[e.g.,][]{Gierliski2008b, Pan2016, Zhang2017}. 
A revised analysis of the QPO detectability in the sample of \citet[][]{Gonzalez2012} with WWZ and LSP 
is necessary and will be presented elsewhere.    
In combination with high Eddington ratios (or small BH masses), the extremely steep 
soft X-ray spectral slope, as presented in Figure 2, would be a useful tool for preselecting promising candidates 
in the search for QPOs among AGNs. 






   

   It has been proposed that HFQPO frequencies scale inversely with the BH masses 
   \citep[$f_{\rm QPO}-M_{\rm BH}$ relation,][]{Remillard2006} and 
   such a relationship covers the entire mass range for all astrophysical BHs, from stellar and
   intermediate-mass to supermassive ones \citep[][]{Zhou2015}. In Figure 3 (upper panel), 
   we reproduce the $f_{\rm QPO}-M_{\rm BH}$ 
   relation by including \src and latest SMBH systems with QPO detections. Clearly, the QPO in \src conforms to the 
   relation within the uncertainty range of $M_{\rm BH}$, strengthening its association with the HFQPOs. 
   The QPO frequency of \src is higher than that found in other AGNs, possibly due to its lower BH mass, which 
   appears to fill the mass gap between AGNs and TDEs. 
   In addition, the QPO frequency does not change for at least 18.5 years. This suggests that the QPO is produced by a 
   stable mechanism and disfavors alternative models that involve temporal processes, such as orbiting 
   hot spots in the disk \citep[e.g.,][]{Schnittman2004}. 
   As we mentioned before, the ultrasoft X-ray spectrum of \src appears to be disk-dominated as in the case of ASASSN-14li. 
   The QPO of J1302 may represent a certain disk oscillation mode rarely seen in other AGNs.  

           \begin{figure}
                   \centering
                 \includegraphics[width=0.45\textwidth]{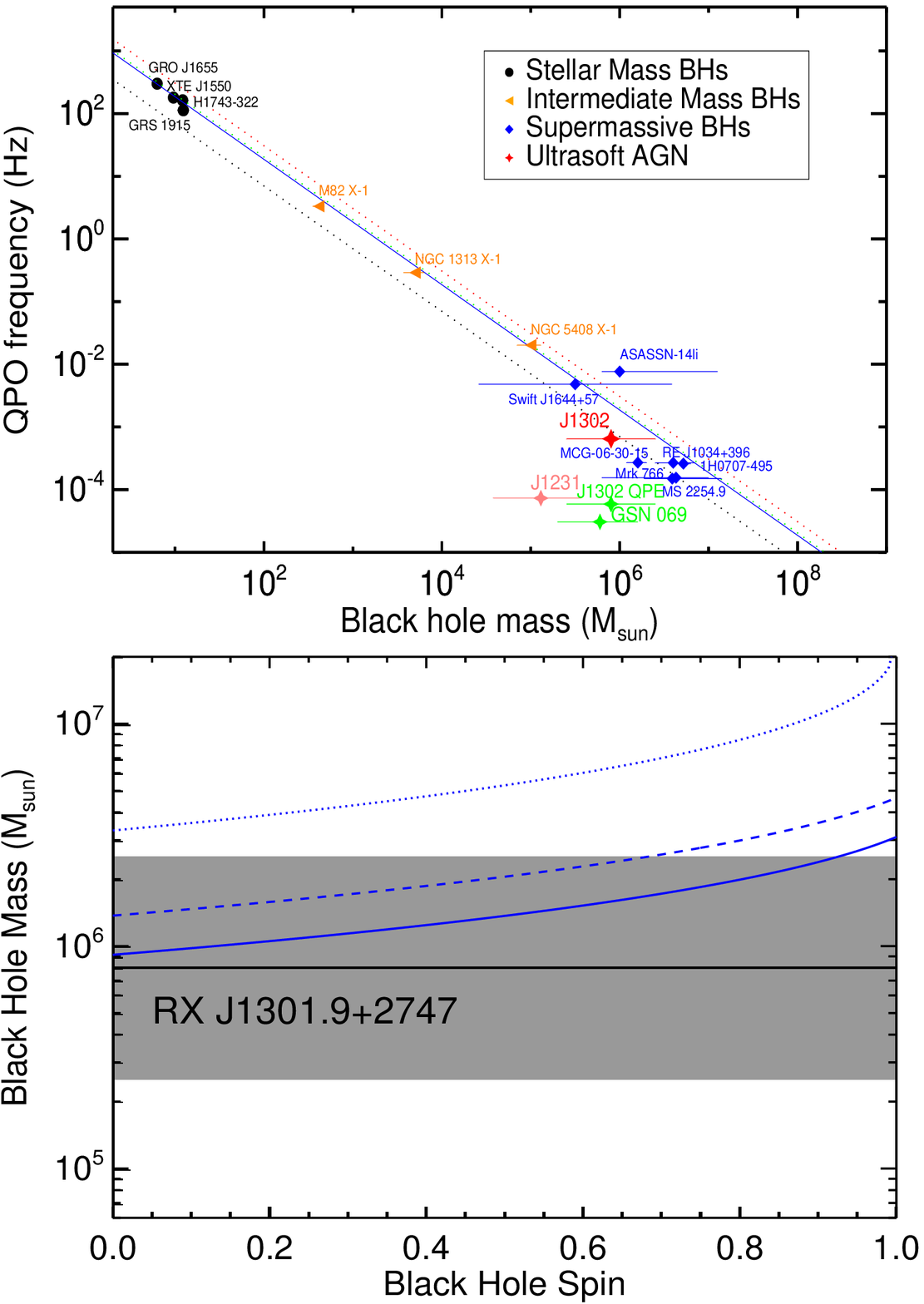}
                    \caption{{\it Upper:} Relation between QPO frequency and BH mass. Following \citet[][]{Pan2016}, we 
                            updated the relation to include more recent QPO detections in NLSy1 MS2254.9-3712 \citep[][]{Alston2015}, 
                            Mrk 766 \citep[][]{Zhang2017}, and MCG -6-30-15 \citep[][]{Gupta2018}, 
                    TDE ASASSN 14li \citep[][]{Pasham2019}, ultrasoft AGN J1231 \citep[][]{Lin2013}, and J1302 (this work). 
                    The two TDEs, ASASSN 14li and Swift J1644+57, have higher QPO frequencies and lower BH masses 
                    in comparison to AGNs. 
                    As a comparison, we 
                    also overplot the frequencies for QPEs \citep[green symbols,][]{Miniutti2019, Giustini2020}. 
                    {\it Lower:} Relation between BH mass and dimensionless spin parameter assuming that the QPO originates 
                    from the {3:2 twin frequency disk resonance} (see text for details). {Only the case for the prograde orbit is shown.}
                    Solid and dashed lines represent cases 
                    in which the observed QPO frequency of J1302 corresponds to the radial and vertical epicyclic frequencies, respectively. 
                    For comparison, we also show the relation by assuming that the QPO frequency is the Keplerian frequency at ISCO (dotted line).                  
                    Grey shaded region shows the J1302's BH mass range ($[0.25-2.5]\times10^6$\msun, Sun13) and the thick black line 
                    represents the best estimate on the BH mass ($8\times10^5$\msun).  
                    }
                        \label{FigGam}%
               \end{figure}

    We attempted to compare our QPO frequency with the three fundamental frequencies for a test particle moving around a spinning BH. 
   The fastest one is the Keplerian orbital frequency ($\nu_{\phi}$). Perturbations can induce two additional frequencies 
   in the radial {or the vertical directions, known as the radial ($\nu_{r}$) or the vertical epicyclic ($\nu_{\theta}$) 
   frequencies,} respectively. 
   According to the simplest version of the 3:2 resonance model as proposed by \citet[][]{Kluzniak2001}, the BH mass and spin relation 
   can be inferred (Figure 3, lower panel) by assuming that the observed QPO frequency corresponds to the radial epicyclic 
   (solid line) and the vertical epicyclic (dashed line) frequencies, respectively. 
   For the case of radial epicyclic oscillation mode, that is, the QPO frequency corresponds to the lower 
   of the twin peak frequencies, our best estimate on the BH mass range ($M_{\rm BH}=[0.5-2.5]\times10^6$\msun) 
   {enables us to constrain a dimensionless spin parameter $a^{*}\simlt0.9$, 
   indicating that a maximal BH spin can be ruled out.}
   Associating the QPO frequency with the higher vertical epicyclic frequency will push the upper limit to 
   a lower spin value of $\sim$0.7. 
   This is somehow consistent with the low BH spin of \src inferred by \citet[][]{Middleton2015} 
   based on the spectral model of Doppler disk tomography. 
   Within the BH mass range, we cannot find formal solutions when choosing 
   the Keplerian frequency at the innermost stable circular orbit (ISCO) for the case of 
   the prograde orbit ($a^{*}>0$). 
   Alternatively, the BH mass could be 
   underestimated, for instance, $M_{\rm BH}\simgt5\times10^6$\msun~is required to conform with the situation of higher 
   frequencies and spin parameters. 
   In this case, however, a lower accretion rate would be expected ($L_{\rm bol}/L_{\rm Edd}<0.02$), 
   which {appears to be at odds} with the disk-dominated thermal spectral state observed for the source.  

   
   
   
   In Figure 3 (upper panel), we also overplot the frequencies for the QPEs, a new phenomenon of X-ray variability 
   associated with accreting SMBHs. As mentioned, while only three ultrasoft AGNs are found (\srcs, J1231, and GSN 069), 
   all of them display quasi-periodic X-ray variations in either form of QPOs or QPEs. We performed similar search for QPO in the quiescent 
   state of GSN 069, 
   but did not find any significant signals. Thus, \src is the only AGN known so far to present both QPOs and QPEs, 
   with the period ratios of $\sim8-13$. Whether there is a physical connection between QPO and QPE is completely unclear. 
   \citet[][]{Miniutti2019} proposed that the observed QPO in J1231 itself may be the signature of a weak QPE or a QPE 
   that has not fully developed. We note that both J1231 and GSN 069 exhibit long-term X-ray decays that can be explained as 
   decade-long sustained TDEs \citep[][]{Lin2017, Shu2018}. It is possible that the transient accretion in the case of TDEs triggers a disk instability, leading to the production of QPO/QPEs as the mass 
   accretion rate drops. This is consistent with the transient nature of QPO/QPEs in both objects, which appear only at later times of 
   the luminosity evolution. Further observations of GSN 069 will be crucial in determining the time evolution of QPEs and 
   testing the possibility that QPEs could develop into a QPO. 
   The extreme variability phenomenon of QPEs is relatively new and we are only beginning to understand 
   their underlying physics. 
   Combined with QPEs, our discovery of the very persistent (stable {over a decade}) QPO in J1302 may present a 
   new probe into the complex accretion physics taking place in an AGN soft state that remains a poorly explored regime. 

\begin{acknowledgements}
 The authors thank the \xmm instrument teams and
 operations staff for making the observations of RX J1301.9+2747 available. 
The work is supported by Chinese NSF through grant Nos. 11822301, 11833007 and U1731104. 
Y.Q.X. acknowledges support from NSFC-11890693, NSFC-11421303, 
the CAS Frontier Science Key Research Program (QYZDJ-SSW-SLH006), and the K.C. Wong Education Foundation. 
C.J. acknowledges support from NSFC-11873054, and the support by the Strategic Pioneer Program on Space Science, CAS, 
through grant Nos. XDA15052100.
\end{acknowledgements}

 \begin{appendix}
         \section{Supplementary figures}
\clearpage       
   \begin{figure*}
           \centering
           \includegraphics[width=0.9\textwidth]{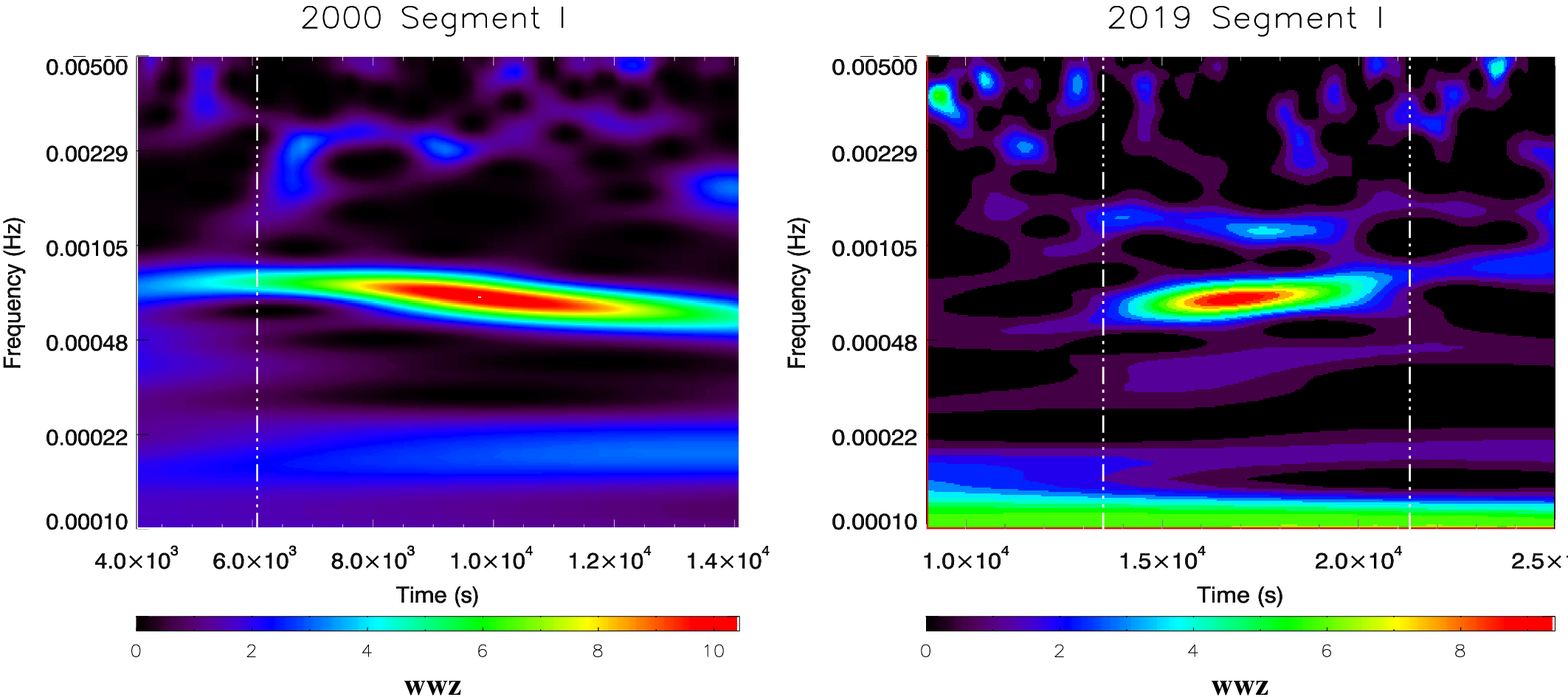}
            \caption{
            Contour plot of the WWZ power for the Segment I of light curve in the quiescent state (see the left panel of Figure 1 in the main text). 
            The WWZ power is plotted in the time-frequency (period) plane 
            (XMM2000 observation on the left and XMM2019 observation on the right). 
            {The color scale of the WWZ power is shown in the lower panel. }
            The significant regions are within a frequency range of $\sim6\times10^{-4}-8\times10^{-4}$Hz for both observations.  
            It can be seen that the QPO signal does not appear for the whole light curve, which is relatively strong in the 
            time segment of 6000-14100 s for the XMM2000 data, and of 13500-21300 s for the XMM2019 data, respectively. 
            The time segments between the vertical dash-dotted lines are selected for the power spectra analysis in the main text. 
            }
                 \label{mdot}%
                 \vspace{1.6cm}          
       \end{figure*}     
         \vspace{0.6cm}

         \begin{figure*}
                 \centering
                 \includegraphics[width=0.9\textwidth]{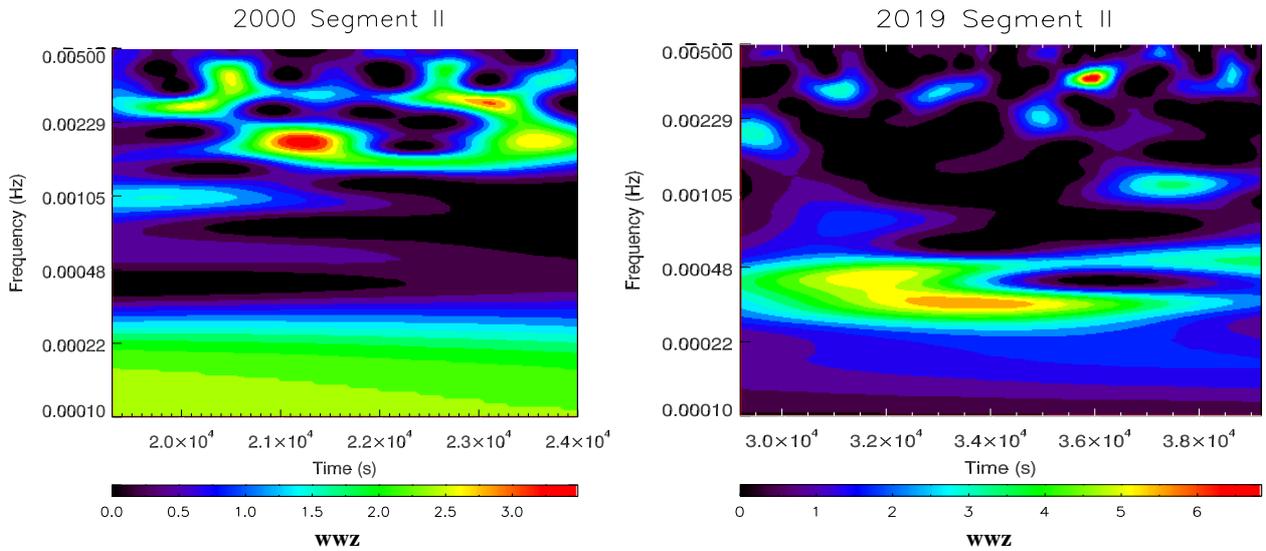}
                  \caption{
            Same contour plot of the WWZ power as Figure A.1., but for segment II of the light curve. 
            {No significant quasi-periodic signal is found}. 
            }
                       \label{mdot}%
             \end{figure*}

         \begin{figure*}
                 \centering
                 \includegraphics[width=0.85\textwidth]{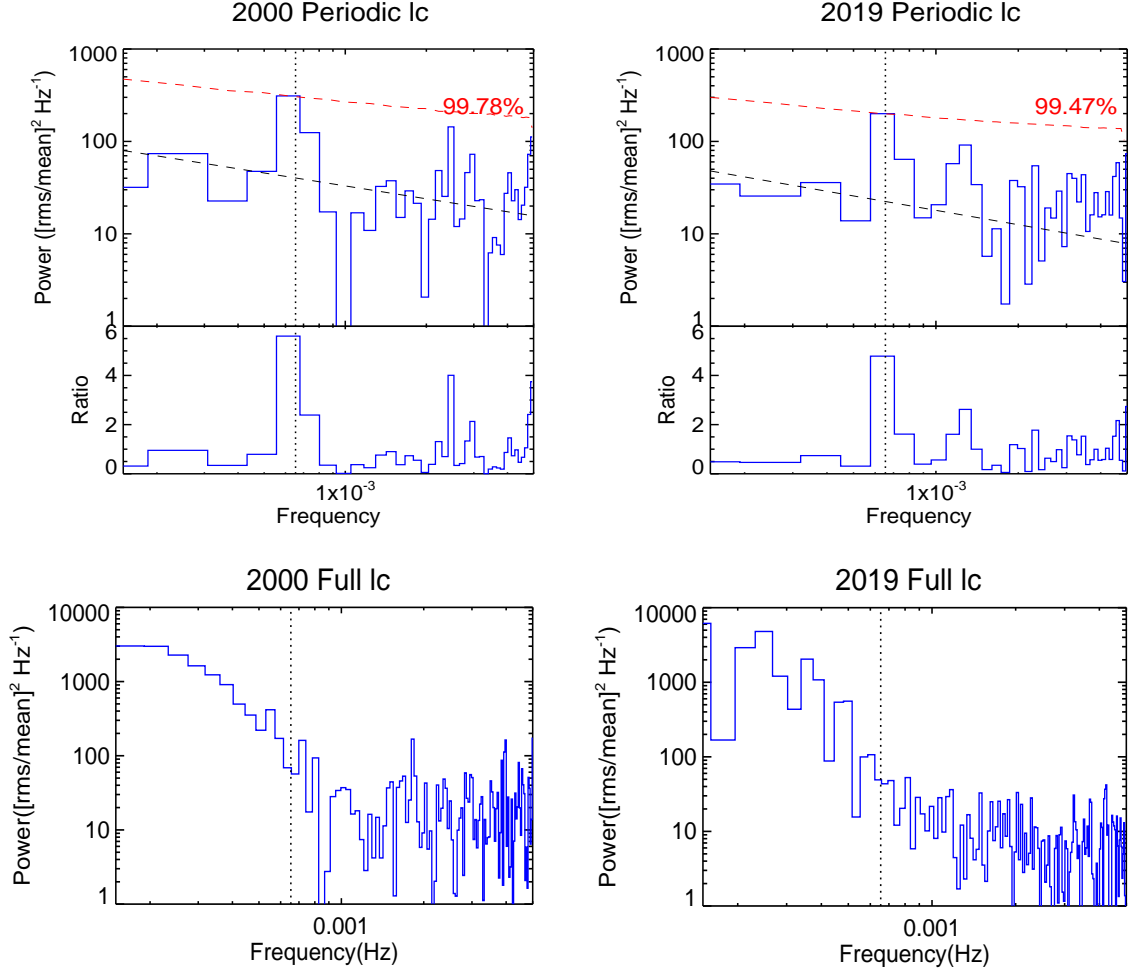}
                  \caption{
          {PSD calculated as the modulus-squared of the DFT, which is normalized with $\rm [rms/mean]^2$. The top panel shows the PSD of light curve from the periodic segment (XMM2000 data on left and XMM2019 data on right). The red dashed line shows the single-trial 99.78\% significance curve, while the black dashed line shows the best-fitted PSD model with a powerlaw plus constant ($ P (f )=N f^{-\alpha}+ C $ ). 
          The corresponding data/model ratios are shown in the lower panel. 
          The vertical line marks the QPO frequency found from the LSP and WWZ analyses (see text). 
          The bottom panel shows the PSD for the full light curve of two XMM observations where no QPO is seen at 
          the frequency of $\sim6.7\times10^{-4}$ Hz.}
            }
                       \label{mdot}%
                 \vspace{0.8cm}          
             \end{figure*}

         \begin{figure*}
                 \centering
                 \includegraphics[width=0.9\textwidth]{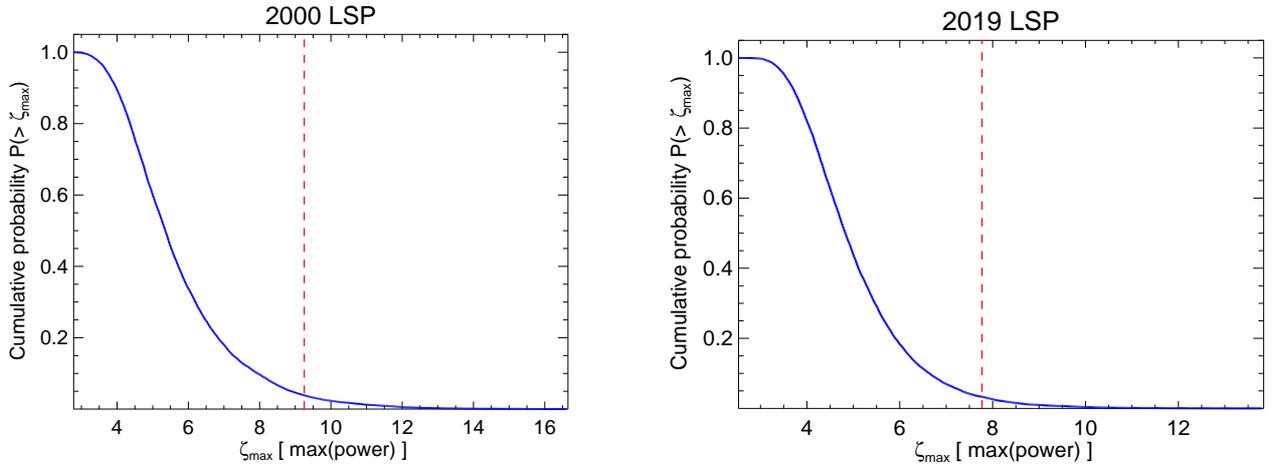}
                  \caption{
         { Left: Cumulative distribution of global chance probability of the maximum LSP power ($\zeta_{\rm max}$) searched in the given frequency range of interest ($1.2\times10^{-4}-5.1\times10^{-3}$ Hz). The probability $P(>\zeta_{\rm max})$ is calculated as the values exceeding a given maximum power. The red vertical line represents the probability of maximum LSP power exceeding the observed QPO value, $\zeta_{\rm max,obs}=9.25$ for the XMM2000 data, which corresponds to a global statistical significance of 96.81\% for the QPO detection. Right: Same as left, but for the simulations of XMM2019 data. }
            }
                       \label{mdot}%
             \end{figure*}

         \end{appendix}     

\end{document}